\documentclass[superscriptaddress,twocolumn]{revtex4}
\usepackage{amssymb}
\usepackage{amsfonts}
\usepackage{amsmath}
\usepackage{graphicx}
\usepackage{dcolumn}
\usepackage{bm}
\usepackage[latin1]{inputenc}
\usepackage[dvips]{hyperref}

\begin{document}

\title{Compact extra dimensions in cosmologies with $f(T)$ structure}

\author{Franco Fiorini}
\email{francof@cab.cnea.gov.ar.}
\affiliation{Centro Atómico Bariloche, Comisión Nacional de Energía Atómica
R8402AGP Bariloche, Argentina.}
\affiliation{Sede Andina, Universidad Nacional de Río Negro, Villegas 137 (8400),
Bariloche, Argentina.}
\author{P. A. González}
\email{pablo.gonzalez@udp.cl}
\affiliation{Facultad de Ingeniería. Universidad Diego Portales, Avenida Ejército
Libertador 441, Casilla 298-V, Santiago, Chile.}
\author{Yerko Vásquez}
\email{yvasquez@ufro.cl}
\affiliation{Departamento de F\'{\i}sica, Facultad de Ciencias, Universidad
de La Serena,\\
Avenida Cisternas 1200, La Serena, Chile.}
\affiliation{Departamento de Ciencias Físicas, Facultad de Ingeniería, Ciencias y
Administración, Universidad de La Frontera, Avenida Francisco Salazar
01145, Casilla 54-D, Temuco, Chile.}

\begin{abstract}
The presence of compact extra dimensions in cosmological scenarios in the context of $f(T)$-like gravities is discussed. For the case of toroidal compactifications, the analysis is performed in an arbitrary number of extra dimensions. Spherical topologies for the extra dimensions are then carefully studied in six and seven spacetime dimensions, where the proper vielbein fields responsible for the parallelization process are found.
\end{abstract}

\maketitle

\section{Introduction}
\label{sec:intro}

In 1928 Einstein proposed an equivalent formulation of General Relativity (GR) nowadays known as the Teleparallel Equivalent of General Relativity (TEGR) \cite{ein28,Hayashi79}. In this theory, the Weitzenb\"{o}ck connection is used instead of the Levi-Civita connection to define the covariant derivative. Weitzenb\"{o}ck connection has non-null torsion; however, it is curvatureless, which implies that this formulation of gravity exhibits only torsion. In $D$ spacetime dimensions, the dynamical fields are the $D$ linearly independent vielbeins and the torsion tensor is formed solely by them and the first derivatives of these objects. The Lagrangian density, which will be noted hereafter as $T$, is constructed from this torsion tensor, assuming invariance under general coordinate transformations and local Lorentz transformations, along with demanding that the Lagrangian density be quadratic in the torsion tensor \cite{Hayashi79}. In recent years an extension of the above Lagrangian density was constructed in \cite{Nos1,Bengochea:2008gz,Nos2,Linder:2010py}, making the Lagrangian density a function of the scalar torsion $T$, the so-called $f(T)$ gravity. Much attention has been focused on this extended $f(T)$ gravity theory recently, because it exhibits interesting cosmological implications, as witnessed in the study of missing matter problems, as well as providing a new mechanism to explain the late acceleration of the universe based on a modification of the gravitational theory, instead of introducing an exotic content of matter (dark energy problem) \cite{Zheng:2010am, Bamba:2010wb, Zhang:2011qp, Capozziello006, Bamba:2011pz, Wei:2011yr, Geng:2011ka, Karami:2012fu, Yang:2012hu, Bamba:2012vg, Xu:2012jf,
Ghosh:2012pg}. It is also believed that $f(T)$ gravity could be a reliable approach to address the shortcomings of general relativity at high-energy scales \cite{review}. For instance, in \cite{Nos2} a Born-Infeld $f(T)$ gravity Lagrangian was used to cure the physically inadmissible divergencies occurring in the Big Bang of standard cosmology, rendering spacetime geodesically complete and powering an inflationary stage without the introduction of an inflaton field.

We are now interested in studying this alternative theory of gravity in a higher dimensional context, and  the basic features of cosmological behavior due to this extension. The conception of extra dimensions in physical theories has a long history that begins with the original ideas of Kaluza and Klein \cite{Kaluza:1921tu} and finds a new realization nowadays in modern string theory \cite{Font:2005td, Chan:2000ms}. One of the main motivations for studying these higher dimensional models is the chance of unifying the fundamental interactions of nature. In most extra-dimensional scenarios, the additional dimensions are assumed to be compactified on a very small (unobservable) internal submanifold, whereas the other spacetime dimensions constitute the 4-dimensional observable universe. Another mechanism of dimensional reduction studied in cosmological GR scenarios is  dynamical compactification \cite{Chodos:1979vk,Freund:1982pg,Dereli:1982ar}. In this case, the internal dimensions evolve in time to very small scales as the external dimensions expand, and so the observable universe becomes effectively 4-dimensional. This type of reduction was first studied by Chodos and Detweiler in \cite{Chodos:1979vk}, who considered vacuum Einstein equations in five spacetime dimensions. These results were latter extended, and a full classification of homogeneous eleven-dimensional cosmologies can be found in \cite{Demianski:1987rm}. Dynamical reduction in multidimensional Bianchi type I models was studied in \cite{Szydlowski} and a study of multidimensional cosmological models as dynamical systems with topology FRW$^{4} \times T^{D-4}$ (here $T^{D-4}$ refers to the $(D-4)$-torus) was performed in \cite{Szydlowski:1990jt, Szydlowski:1990ju}. More recently, and in a totally  different context, GR in the presence of a large number of extra dimensions was carefully studied in \cite{emparan:2013, Gaston:2013}.

The specific topic of extra dimensions have scarcely been worked out in $f(T)$ theories, and it remits, in one hand, to some black hole solutions of $f(T)$ gravity in arbitrary spacetime dimensions \cite{Capozziello:2012zj}, and on the other hand, to an analysis of four-dimensional $f(T)$ gravities as effective theories coming from five-dimensional Kaluza-Klein and Randall-Sundrum models \cite{Odintsov2}. In this work we investigate for the first time cosmological solutions in the context of $f(T)$ gravity in higher dimensions, considering that the extra dimensions are compactified in different ways. This investigation is motivated primarily by the increasing interest in multidimensional cosmological scenarios, particularly brane-world models \cite{maartens:2010}. In the present work we shall consider both toroidal and spherical topologies for the internal dimensions. With this purpose in mind we proceed first to obtain the proper parallel vector fields for the different topologies under consideration. This topic is a difficult one, and constitutes the starting point of any model describing extra dimensions in gravitational theories with absolute parallelism, $f(T)$ gravity being one of them. In particular we consider five, six, and seven-dimensional models and study all the possible compactifications of the extra dimensions using the right vielbein field that parallelizes the spacetime in each of these cases, and we then proceed to obtain the corresponding field equations. Then we show in general terms, that in some limited cases, simple analytical solutions describe physically relevant spacetimes representing different stages of cosmic evolution. The results presented here are mandatory aspects required to deal with more fundamental topics of physical cosmology such as structure formation and cosmological perturbations.

The paper is organized as follows: in section~\ref{sec:fundamental} we present a brief review
of the fundamentals of $f(T)$ theories. In section~\ref{sec:toroidal} we study toroidal
compactification for an arbitrary number of extra dimensions, and in section~\ref{sec:spherical} we study spherical
compactifications with the emphasis on six and seven spacetime dimensions. Finally, in section~\ref{sec:conclution} we present the conclusions.

\section{Fundamentals of $f(T)$ theories}
\label{sec:fundamental}
The extended gravitational schemes with absolute parallelism ($f(T)$
theories), take as a starting point the Teleparallel Equivalent of General
Relativity. We will summarize here the basic elements needed to elaborate the ideas of the present work, leaving the details for
the reader. For a thorough introduction to $f(T)$ gravity as well as to its mathematical basis the reader can consult for instance \cite%
{Nos3,Nos4, Izumi, Barrow, Youssef}.

\bigskip

The spirit of the equivalence between the Riemann and Weitzenb\"{o}ck
formulations of GR can be summarized in the equation
\begin{equation}
T\, =\, -R + 2\; e^{-1}\;\partial _{\nu }(e\,T_{\sigma }^{\ \sigma \nu
}\,)\;.  \label{divergence}
\end{equation}
On the left hand side of Eq. (\ref{divergence}) we have the so-called
Weitzenb\"{o}ck invariant
\begin{equation}
T\ =\ S_{\ \mu \nu }^{\rho }\ T_{\rho }^{\ \mu \nu }\, ,  \label{Weitinvar}
\end{equation}
where $T_{\rho }^{\,\,\,\mu \nu}$ are the components (on a coordinate basis) of the torsion two-form
$T^{a}=de^{a}$ coming from the Weitzenb\"{o}ck connection $\Gamma^\lambda_{\nu%
\mu}=\,e_{a}^{\lambda }\,\partial _{\nu }e_{\mu }^{a}$, and $S_{\lambda \mu
\rho }$ is defined according to
\begin{equation}
S_{\ \mu \nu }^{\rho }=\frac{1}{4}\,(T_{\ \mu \nu }^{\rho }-T_{\mu \nu }^{\
\ \ \rho }+T_{\nu \mu }^{\ \ \ \rho })+\frac{1}{2}\ \delta _{\mu }^{\rho }\
T_{\sigma \nu }^{\ \ \ \sigma }-\frac{1}{2}\ \delta _{\nu }^{\rho }\
T_{\sigma \mu }^{\ \ \,\sigma }.  \label{tensor}
\end{equation}
Actually, $T$ is the result of a very specific quadratic combination of
irreducible representations of the torsion tensor under the Lorentz group $%
SO(1,3)$ \cite{Hehl}. Equation \eqref{divergence} simply says that the
Weitzenb\"{o}ck invariant $T$ differs from the scalar curvature $R$ in a total
derivative; therefore, both conceptual frameworks are totally equivalent at
the time of describing the dynamics of the gravitational field.

$f(T)$ gravity can be viewed as a natural extension of Einstein gravity, and
is ruled by the action in $D$ spacetime dimensions
\begin{equation}
S=\frac{1}{16 \pi G }\int d^{D}x\,e\left[ f(T) +L_{matter}\right] ~,
\label{actionfT3D0}
\end{equation}%
being $e=\sqrt{det(g_{\mu\nu})}$. Of course GR is contained in (\ref{actionfT3D0}) as the particular case when $f(T)=T$. The dynamical
equations in $f(T)$ theories, for matter coupled to the metric in the usual
way, are
\begin{eqnarray}
 &&\left(e^{-1}\ \partial_\mu(e\ S_a^{\ \ \mu\nu})\, +\, e_a^\lambda \
T^\rho_{\ \ \mu\lambda}\ S_\rho^{\ \ \mu\nu}\right)\, f^{\prime }(T)\, + \notag \\
&& S_a^{\ \ \mu\nu}\ \partial_\mu(T)\ f^{\prime \prime }(T)\, -\frac{1}{4}
e_a^\nu\ f(T)\ =\ -4\pi G\ e_a^\lambda\ T_\lambda^{\ \nu}\ , \label{ecuaciones}
\end{eqnarray}
where the prime means derivative with respect to $T$ and $T_\lambda^{\ \nu}$ is the energy-momentum tensor. It is worth
insisting that equations \eqref{ecuaciones} are second-order
differential equations for the vielbein components. This implies a genuine
advantage compared with other deformed gravitational schemes such as, for
instance, the popular $f(R)$ gravities. This goodness, however, has
consequences: the lack of local Lorentz invariance. If we perform on a given
solution (let us say $e^{b}$), a local Lorentz transformation $%
e^{a}\longrightarrow e^{a^{\prime}} =\Lambda _{b}^{a^{\prime}}(x)\ e^{b}$,
we will find that $e^{a^{\prime}}$ is not a solution of the field equations
\eqref{ecuaciones}. The reason for this lack of local invariance is simple: under a local transformation the scalar \eqref{Weitinvar} changes according to $T\rightarrow T'=T+$ surface term. So this surface term, which is harmless when $f(T)=T$, remains inside the function $f$, thus ruining the local invariance of the theory. This is because the theory picks up preferred
referential frames that constitute the autoparallel curves of the given
manifold \cite{christian}. In other words, the field equations \eqref{ecuaciones} determine
the full components of the vielbein and not just those of the metric tensor,
related to the vielbein by means of
\begin{equation}
g_{\mu \nu }(x)=\eta _{ab}\ e_{\mu }^{a}(x)\ e_{\nu }^{b}(x).
\label{metric}
\end{equation}
The preferred reference frames that solve the dynamical equations \eqref{ecuaciones} constitute what we shall call the \emph{proper frames}. This frames define the spacetime structure by means of a global set of basis covering the whole tangent bundle $\mathcal{T}%
(\mathcal{M})$, i.e., they \emph{parallelize} the spacetime (see section \ref{sec:spherical} for details). From the infinite set of vielbeins giving a certain metric, the motion equations \eqref{ecuaciones} choose those frames which act as well defined (smooth), non null fields on the whole tangent bundle. Equation (\ref{metric}) just says that these smooth basis fields are everywhere orthonormal.

An important fact sometimes omitted in the literature concerns the coupling with the matter fields. If we assume that the matter action depends only on the metric (and not on the whole vielbein), then the non-fermionic matter fields are unable to sense the lack of local Lorentz invariance by virtue of the invariant character of expression \eqref{metric} under such transformations. Then, as far as coupling to (non-spinning) matter is concerned, the lack of local invariance should not be problematic. The effects of the additional degrees of freedom that certainly exist in $f(T)$ theories as a consequence of the breaking of the Lorentz invariance (see the analysis performed in \cite{Miao:2011}), must be found beyond the unspinning matter. Actually, it was found that on the flat FRW background with a scalar field, linear perturbation up to second order does not reveal any extra degree of freedom at all \cite{Izumi}. It is fair to say, though, that the nature of the additional degrees of freedom remains unknown.

 \bigskip

\section{Toroidal compactifications in $D-4$ internal dimensions}
\label{sec:toroidal}
In this section we will study a $D$-dimensional cosmological model inside $f(T)$-gravity theory. We will consider the simplest case where the
additional internal dimensions are compactified in a torus, and in the
next section we will discuss other topologies for the extra dimensions. The external spacetime is assumed to be spatially flat in accordance with the current experimental evidence. Thus, the topology of our cosmological model is described by $R\times
R^{3}\times S^{1}\times ...\times S^{1}$, where $S^{1}\times ...\times S^{1}$
is the $D-4$ dimensional torus. Under these considerations the
metric is written as
\begin{equation}
ds^2=dt^2-a_{0}^2(t)(dx^2+dy^2+dz^2)-\sum_{n=1}^{D-4}a_{n}(t)\,dX_{n}^2 ,
\label{metD}
\end{equation}
where $a_{0}(t)$ denotes the scale factor of the external space and $a_{n}(t)$ ($n=1,...,D-4$) the scale factors for the internal
dimensions $X_{n}$. Based on these observations we know that the internal compact dimensions must be very small in order to be undetectable, so we are mainly interested in solutions where all the radii of the internal
dimensions $a_{n}(t)$ ($n=1,...,D-4$) become smaller as cosmic time evolves. To proceed, we first need to find the correct parallelization of the spacetime under consideration. For the metric \eqref{metD}, and due to
the fact that the circle $S^{1}$ is trivially parallelizable (just take any smooth, non null tangent vector field in $\mathcal{T}(S^1)$), a vielbein that
parallelizes spacetime is a simple diagonal one
\begin{equation}
e^a_{\mu}=diag\left(1,a_{0}(t),a_{0}(t),a_{0}(t),a_{1}(t),...,a_{D-4}(t)%
\right).  \label{vielD}
\end{equation}
Actually, we can think about the vielbein (\ref{vielD}) as follows: unwrap the periodic coordinates $X_{n}$, and take the submanifold $t=constant$. After rescaling the coordinates according to

\begin{eqnarray*}
&&x\rightarrow x'=a_{0}\,x \\
&&y\rightarrow y'=a_{0}\,y \\
&&z\rightarrow z'=a_{0}\,z \\
&&X_{n}\rightarrow X'_{n}=a_{n}\,X_{n},
\end{eqnarray*}
we just get $D-1$ Euclidian space. For this space, it is clear that the autoparallel lines are just the straight lines which can be generated by the basis $\partial_{i}$ $(i:1,...D-1)$, the dual cobasis of which is $dx_{i}$. So, up to a time-dependent conformal factor, the frames describing the autoparallel lines are $dx_{i}$, and then the whole spacetime is parallelized as it is indicated in \eqref{vielD}.

In order to obtain the motion equations, we shall work in the co-moving system, where the energy-momentum tensor for a perfect fluid reads
\begin{equation}
T^{\mu}_{\nu}=diag(\rho,-p_{0},-p_{0},-p_{0},-p_{1},...,-p_{D-4}).
\label{TeneDos}
\end{equation}
In this way, the energy conservation equation takes the form
\begin{equation}
\dot{\rho}+(3H_{0}+\sum_{n=1}^{D-4}H_{n})\rho+3H_0p_0+%
\sum_{n=1}^{D-4}H_np_n=0,  \label{EcconD}
\end{equation}
where a dot means derivative with respect to time and $H_{0}=\dot{a}_{0}/a_{0}$ and $H_{n}=\dot{a}_{n}/a_{n}$ are the
Hubble parameters of the submanifolds with the scale factors $a_{0}(t)$ and $%
a_{n}(t)$, respectively. We can now compute the Weitzenb\"{o}ck invariant \eqref{Weitinvar}, which is
\begin{equation}
T=-6\left( H_{0}^{2}+H_{0}\,\sum_{n=1}^{D-4}H_{n}+\frac{1}{3}%
\sum_{n=1}^{D-4}H_{n}\,H_{n+1}\right).  \label{inv6deA}
\end{equation}
The initial value equation, i.e., the equation coming from the variation of the action respect the $e^0_{0}$ component of the vielbein results
\begin{equation}
f-2 \,T f^{\prime }=16 \pi G \rho.  \label{eqvi6d}
\end{equation}
On the other hand, the action variation respect the spatial sector of the
vielbein, leads to the following equations
\begin{eqnarray}
&&2f^{\prime }\left( 3H_{0}^{2}+2\dot{H}_{0}+\sum_{n=1}^{D-4}(\dot{H}%
_{n}+2H_{0}H_{n}+H_{n}^{2})-\frac{T}{2}\right) +\notag\\
&&2f^{\prime \prime }\dot{T}%
\left(2H_{0}+\sum_{n=1}^{D-4}H_{n}\right)+
f(T) =-16\pi Gp_{0},
\label{eqesp1}
\end{eqnarray}
\begin{eqnarray}
&&2f^{\prime}\left(6H_{0}^{2}+3\dot{H}_{0}+\sum_{n=1,n\neq b}^{D-4}(\dot{H}%
_{n}+H_{0}H_{n}+H_{n}^{2})-\frac{T}{2}\right) +\notag\\
&&2f^{\prime \prime}\dot{T}%
\left(3H_{0}+\sum_{n=1,n\neq b}^{D-4}H_{n}\right)+
f(T) =-16\pi Gp_{b}.
\label{eqesp2}
\end{eqnarray}
Note that this last expression actually contains $D-4$ equations. The full system of equations to be solved is given by Eqs. \eqref{eqvi6d}-\eqref{eqesp2} with $T$ given by \eqref{inv6deA}. They contain the conservation equation \eqref{EcconD} self-consistently.

\bigskip

Now we will investigate some simple limit cases, in order to analyze the behavior
of the scale factors. In all of this section, we will consider that the Hubble parameters of the
extra dimensions are equal ($H_{1}=...=H_{D-4}\equiv H_{1}$) and the content of matter in the spacetime is assumed to be dust
matter, i.e. $p_0=p_n=0$. In this way, Eqs. \eqref{inv6deA}, \eqref{eqesp1},
and (\ref{eqesp2}) are simplified to
\begin{equation}  \label{CTD}
T=-6H_{0}^{2}-6\left( D-4\right) H_{0}H_{1}-\left( D-4\right) \left(
D-5\right) H_{1}^{2},
\end{equation}
\begin{eqnarray}
&&2f^{\prime }\left( 3H_{0}^{2}+2\dot{H}_{0}+(D-4)(\dot{H}%
_{1}+2H_{0}H_{1}+H_{1}^{2})-\frac{T}{2}\right) +\notag\\
&&2f^{\prime \prime }\dot{T}%
\left(2H_{0}+(D-4)H_{1}\right)+f(T) =0,
\end{eqnarray}
\begin{eqnarray}
&&2f^{\prime}\left(6H_{0}^{2}+3\dot{H}_{0}+(D-5) (\dot{H}%
_{1}+H_{0}H_{1}+H_{1}^{2})-\frac{T}{2}\right) +\notag\\
&&2f^{\prime \prime}\dot{T}%
\left(3H_{0}+(D-5)H_{1}\right)+f(T) =0.
\end{eqnarray}
First, we will analyze the limit case when both Hubble parameters, $H_0$ and $H_1$, are constants; therefore, from Eq. \eqref{CTD} we know that $T$ must be a constant too, and the above
equations reduce to a more manageable form
\begin{equation}
2f^{\prime }\left( 3H_{0}^{2}+(D-4)(2H_{0}H_{1}+H_{1}^{2})-\frac{T}{2}%
\right)+f(T) =0, \label{eqesp1bis}
\end{equation}
\begin{equation}
2f^{\prime}\left(6H_{0}^{2}+(D-5) (H_{0}H_{1}+H_{1}^{2})-\frac{T}{2}%
\right)+f(T) =0. \label{eqesp2bis}
\end{equation}
Therefore, the equations
\eqref{eqesp1bis} and \eqref{eqesp2bis} yield the following condition
\begin{equation}  \label{H0D}
H_1=-\frac{6H_0}{\sqrt{(D-3)^2+12}-(D-3)}.
\end{equation}
Having imposed the constancy of $H_{0}$ and $H_{1}$, we will see under what conditions the initial value and energy
conservation equations are satisfied. In this case, the energy conservation
equation can be written as
\begin{equation}  \label{ic}
\dot{\rho}+(3H_{0}+(D-4)H_{1})\rho=0 .
\end{equation}
Using the fact that $H_0=\dot{a}_0(t)/a_0(t)$ and $H_1=\dot{a}_1(t)/a_1(t)$, the energy density is integrated straightforwardly from the above expression, leading to
\begin{equation}  \label{ecd}
\rho=\frac{C}{a_0^3(t)a_1^{D-4}(t)},
\end{equation}
where $C$ is an integration constant.
The initial value equation \eqref{eqvi6d} is satisfied if $\rho $ is
constant (due to the constancy of $T$), which implies that the term $a_{0}^{3}\left( t\right) a_{1}^{D-4}\left(
t\right) $ must be constant. This condition actually means that the total volume of the higher dimensional space remains constant in time, which will ensure that a dynamical contraction of the internal dimensions yields an expanding external space. From Eq. \eqref{ic} we obtain
\begin{equation}
H_{0}=-\frac{\left( D-4\right) }{3}H_{1}.  \label{constraint}
\end{equation}
This equation together with Eq. \eqref{H0D} are only satisfied for $D=5$. However, in five dimensions we have $H_0=-H_1/3$ and $T=12H_0^2$. Thus, by replacing in Eq. \eqref{eqesp2bis}, we obtain $f(T)=0$. In this way, Eq. \eqref{eqvi6d} yields $\rho=0$ and therefore the constant $C$ in \eqref{ecd} is null. Also, it is easy to see that for dimensions higher than five and for $p_0=p_n=\rho=0$ and a constant $H_0$, the only possible solution is given by $H_0=H_1=0$ with $f(T)=0$. Therefore, we conclude that this compactification is not able to describe a de Sitter expansion in vacuum.

\bigskip

Another limit case worth mentioning is when the Hubble parameters of the extra dimensions vanish ($%
H_{1}=...=H_{D-4}=0$), which means that the internal space has a constant volume that should be very small. This represents a very simple model describing the asymptotic evolution of the additional dimensions to a final constant volume. It is easy to see that no solution of such type exists in this context. In contrast, we will see in section \ref{sec:spherical} that this kind of behavior is possible for the extra dimensions in some spherical compactifications.

Finally, let's briefly comment on some additional exact solutions occurring in $D=5$. Taking $\rho=0$ and non-constant Hubble parameters in Eqs.
\eqref{inv6deA}, \eqref{eqesp1}, and \eqref{eqesp2} we get
\begin{equation}
T=-6\left( H_{0}^{2}+H_{0}H_{1}\right)\label{escalar5de},
\end{equation}
\begin{eqnarray}
&&2f^{\prime }\left( 3H_{0}^{2}+2\dot{H}_{0}+\dot{H}_{1}+2H_{0}H_{1}+H_{1}^{2}-%
\frac{T}{2}\right) +\notag\\
&&2f^{\prime \prime }\dot{T}\left(2H_{0}+H_{1}\right)+f(T)
=0,
\end{eqnarray}
\begin{equation}
2f^{\prime}\left(6H_{0}^{2}+3\dot{H}_{0}-\frac{T}{2}\right) +6f^{\prime
\prime}\dot{T}H_{0}+f(T)=0.  \label{eqesp25d}
\end{equation}
From Eq. (\ref{eqvi6d}), we obtain
\begin{equation}
\frac{f}{2f\prime}=T.\label{valinvacio}
\end{equation}
Therefore, $T$ is a constant that depends on $f(T)$ \footnote{Note that if a given function $f(T)$ fulfills Eqn. (\ref{valinvacio}), then $\widetilde{f}(T)=f(T)+\epsilon \sqrt{T}$, with a constant $\epsilon$, also solves the equation. Then, the addition of the squared root term can always be dismissed in vacuum.}, and by replacing it in Eq.
\eqref{eqesp25d}, we get
\begin{equation}
\dot H_0+2H_0^2+\frac{T}{6}=0,\,\,\,\,\,\,\,\ H_1=-\frac{T}{6H_0}-H_0,\label{trescasos}
\end{equation}
where we have used \eqref{escalar5de}. By this manner we can distinguish three cases that depend on the sign of $T$.
\begin{itemize}
\item{Case $T<0$.} Eq. \eqref{trescasos} tells us in this case that the Hubble parameter $H_0$ yields
\begin{equation}
H_0=\frac{1}{2}\sqrt{\frac{\left\vert T\right\vert}{3}}\,tanh\left(\sqrt{\frac{\left\vert T\right\vert}{3}}%
(t+B)\right),
\end{equation}
where $B$ is an integration constant. It is worth noting that both scale factors, which are given by
\begin{equation}
a_{0}\left( t\right) =\alpha \cosh ^{1/2}\left( \sqrt{\frac{\left\vert
T\right\vert }{3}}\left( t+B\right) \right),
\end{equation}
\begin{eqnarray}
a_{1}\left( t\right) &=&\beta \sinh \left( \sqrt{\frac{\left\vert T\right\vert
}{3}}\left( t+B\right) \right)\, \times \notag \\ &&\cosh ^{-1/2}\left( \sqrt{\frac{\left\vert
T\right\vert }{3}}\left( t+B\right) \right),
\end{eqnarray}
where $\alpha $ and $\beta $ are constant, increase in time. This case, thus, seems to be unprovided of physical significance.
 \item	{Case $T>0$.}
Now $H_{0}$ is given by
\begin{equation}
H_{0}=-\frac{1}{2}\sqrt{\frac{T}{3}}\tan \left( \sqrt{\frac{T}{3}}\left(
t+B\right) \right),
\end{equation}
and the scale factors by
\begin{equation}
a_{0}\left( t\right) =\alpha \cos ^{1/2}\left( \sqrt{\frac{T}{3}}\left(
t+B\right) \right),
\end{equation}

\begin{eqnarray}
a_{1}\left( t\right)&=&\beta \sin \left( \sqrt{\frac{T}{3}}\left( t+B\right)
\right)\, \times \notag \\ &&\cos ^{-1/2}\left( \sqrt{\frac{T}{3}}\left( t+B\right) \right) .
\end{eqnarray}
In order to be physically admissible, at least as a model of the early stages of cosmic evolution, we must bind the proper time to the interval $-\frac{\pi}{2}\sqrt{\frac{3}{T}}\leq t\leq 0$, having taken $B=0$. In this way $a_{0}$ is an increasing function of time, while $a_{1}$ decreases from infinite at $t=-\frac{\pi}{2}\sqrt{\frac{3}{T}}$ to a null value at $t=0$.
\item{Case $T=0$.}
Finally, when the scalar torsion vanishes the scale factors are given by
\begin{equation}
a_{0}^{2}\left( t\right) =2t+B,
\end{equation}
\begin{equation}
a_{1}^{2}\left( t\right) =\frac{1}{2t+B}.
\end{equation}
This behavior for the scale factors shows a suppression of the extra dimensions as the bulk expands linearly in time, and then, constitutes a physical admissible simple model of the early universe.
\end{itemize}
Note that in the three cases, the solutions are independent of the specific form of the function $f(T)$, in the sense that the sole role of $f(T)$ is to fix the constant value of $T$. We actually included them here as simple examples of exact states of the theory, and we hope that they might serve as a seed for more realistic solutions.

\bigskip

\section{Spherical compactifications in $D-4$ internal dimensions}
\label{sec:spherical}

\subsection{Statement of the problem}

In theories relying on absolute parallelism, i.e., in theories where the whole vielbein components (not just the metric tensor) are determined by the field equations, the structure of the parallel
vector fields describing the additional compact dimensions is highly non-trivial. In this section we investigate this issue further and discuss a
number of topics essential to understanding the nature of
the additional compact dimensions when they are other than the simple
toroidal compactifications previously discussed. In particular, we shall now focus
on extra compact dimensions with topology $S^{n}$, $n>1$.

Our task is to find cosmological solutions of the $f(T)$ field equations (\ref{ecuaciones}) with $D-4$ spherically compactified extra dimensions, i.e. a set of $D$ one forms $e^a(x)$ that solve the equations and lead to the metric tensor
\begin{eqnarray}
ds^{2}&=&dt^{2}-a_{1}^{2}(t)d\Omega_{(j)}^{2}-a_{2}^{2}(t)d\Omega_{(k)}^{2}-...-\\ \notag
&&-a_{0}^{2}(t)(dx^{2}+dy^{2}+dz^{2}),  \label{metgenesf}
\end{eqnarray}%
where $d\Omega_{(j)}^{2}$ refers to the line element corresponding to the $j$ sphere, so $j+k+...=D-4$. In order to do so, we need first to propose a proper frame field for the topology $R\times S^{j}\times S^{k}\times...\times R^3$, i.e., $D$ one forms $e^a(x)$ which turn equations (\ref{ecuaciones}) in a set of consistent equations for the unknown (time dependent) scale factors $a_{0},a_{1},a_{2},...$. Once this is done, it will remain to solve the equations and to obtain the functional form of the scale factors. In section \ref{sec:toroidal} we have explained how to do this for extra dimensions with toroidal topology, where the proper frame field was given in (\ref{vielD}). Now we want to discuss this issue when the extra dimensions are spherically compactified.

We start noting that a remarkable result due to Stiefel states that every
orientable three-dimensional manifold is also parallelizable \cite{Sti}, thus, so it is $S^{3}$. This means that in every orientable three-manifold $\mathcal{M}_{3}$ there exist three independent smooth, non null vector fields covering the entire tangent bundle $\mathcal{T}(\mathcal{M}_{3})$. Additionally, Kervaire and Milnor \cite{Ker} have shown that, apart from $S^{1}$ and $S^{3}$, the only other parallelizable sphere is $S^{7}$. For our purposes
this means that extra dimension spherical compactifications other than these
will possess a rather complicated parallel vector field structure.

In general, a $D$-dimensional manifold $\mathcal{M}_{D}$ will be parallelizable if a global basis exist in $\mathcal{T}(\mathcal{M}_{D})$. This global basis of the tangent bundle constitutes a preferred reference frame which can be used for defining the space structure. This is so because in this last case we can define a given spacetime as the pair $(\mathcal{T}(\mathcal{M}_{D}),e^a(x))$ instead of $(\mathcal{M}_{D},g_{\mu\nu}(x))$. This preferred reference frame $e^a(x)$ and the solution of the equations (\ref{ecuaciones}), as we shall see, are closely related.

Remaining ever conscious of the subtleties involved, and in order to illustrate the problem, let us consider first two spherically compactified extra dimensions. In this case we need to find a parallelization $e^a(x)$ for the spacetime having the metric
\begin{equation}
ds^{2}=dt^{2}-a_{1}^{2}(t)(d\theta ^{2}+\sin ^{2}\theta \ d\phi
^{2})-a_{0}^{2}(t)(dx^{2}+dy^{2}+dz^{2}),  \label{metD=6}
\end{equation}%
where $(\theta ,\phi )$ are standard spherical coordinates on the 2-sphere.
It is intuitively obvious that the $6-$dimensional manifold described by (%
\ref{metD=6}) is parallelizable. As a matter of fact, the spacetime
topology is $\mathcal{M}_{3}\times R^{3}$, where $\mathcal{M}_{3}\approx
R\times S^{2}$, so the full spacetime manifold can be written as a product
of two $3-$dimensional, orientable (i.e., parallelizable) manifolds.
However, it is important to realize that the parallel vector fields of $%
\mathcal{M}_{3}$ will be highly non-trivial because $S^{2}$ itself is not
parallelizable due to the hairy ball theorem, which states that $S^2$ (like all the even spheres) does not admit a global basis of the tangent bundle $\mathcal{T}(S^{2})$ (see, for instance \cite{Hai}). In order to better understand this point, Figure \ref{esferas} below show two attempts of covering $\mathcal{T}(S^{2})$ with a global, smooth, non null vector field. No matter what the arrangement of the field be, it will fail to be non null or smooth in at least one point of $\mathcal{T}(S^{2})$.

\begin{figure}[h]
\begin{center}
\includegraphics[width=0.35\textwidth]{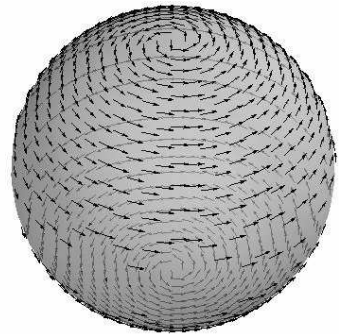}
\includegraphics[width=0.35\textwidth]{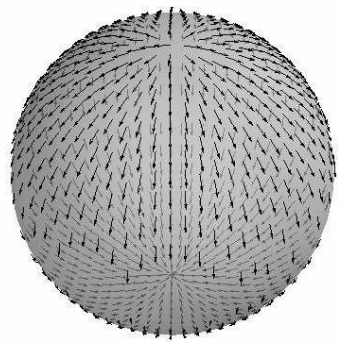}
\end{center}
\caption{Two infructuous attempts of parallelizing $S^2$. In the upper panel the field is smooth but null at the poles. In the lower panel it is non null at the poles but instead it fails to be smooth there.}
\label{esferas}
\end{figure}

Taking into account these facts, we can arrange the components of the vielbein corresponding to the metric (\ref{metD=6}) in a $6\times 6$
matrix $e_{\mu }^{a}$, so schematically we have
\begin{equation}
e_{\mu }^{a}=\left(
\begin{array}{cccccc}
&  & \vline &  &  &  \\
& \mathbb{M}_{3} & \vline &  & \mathbb{O} &  \\
&  & \vline &  &  &  \\ \hline
&  & \vline & a_{0} & 0 & 0 \\
& \mathbb{O} & \vline & 0 & a_{0} & 0 \\
&  & \vline & 0 & 0 & a_{0} \\
&  &  &  &  &
\end{array}%
\right)   \label{esquema}
\end{equation}%
where $\mathbb{O}$ is the $3\times 3$ null matrix and $\mathbb{M}_{3}$ is
the $3\times 3$ matrix representing the parallel \emph{triad} field of $%
\mathcal{M}_{3}$. The main point we wish to emphasize is that $\mathbb{M}_{3}
$ \emph{cannot} be diagonal. If this were so, it would mean that the global
vector fields which perform the paralellization of $\mathcal{M}_{3}$ are
just $e^{1}=dt$, $e^{2}=a_{1}(t)d\theta $ and $e^{2}=a_{1}(t)\sin \theta \,d\phi$, but this is impossible because in this case the fields $e^{2}$ and $e^{3}$ should be a parallelization of $S^{2}$, which does not exist at all.

Even more, it is clear that $\mathbb{M}_{3}$ cannot have the
\emph{block} form
\begin{equation}
\mathbb{M}_{3}=\left(
\begin{array}{ccc}
1\,\,\vline & 0 & 0 \\ \hline
0\,\,\vline &  &  \\
0\,\,\vline & \,\,\,\mathbb{M}_{2} &
\end{array}%
\right) ,  \label{bloque}
\end{equation}%
either, where $\mathbb{M}_{2}$ is some $2\times 2$ matrix. In this case, we
would have that $\mathbb{M}_{2}$ is related to the matrix $%
a_{1}(t)diag(d\theta ,\sin \theta \,d\phi )$ by means of a local rotation.
However, again, it is impossible to find a global basis for $\mathcal{T}%
(S^{2})$. This means that the matrix $\mathbb{%
M}_{3}$ should be related to $diag(dt,a_{1}(t)d\theta ,a_{1}(t)\sin \theta
\,d\phi )$ by a Lorentz boost, and therefore the structure \eqref{bloque} can not be correct. Actually, we should mention that the proper (autoparallel)
vector fields of the form \eqref{esquema} corresponding to the metric \eqref{metD=6} are
unknown at present, and the task of finding the parallelization of $%
\mathcal{M}_{3}$ remains as an open problem. Fortunately, a parallelization of the manifold \eqref{metD=6} can be found in a somewhat different manner, and this will be matter for the paragraphs below. There we will find that the correct vector fields have the structure
\begin{equation}
e^a_{\mu}=\left(
\begin{array}{ccc}
1\,\,\vline & 0 & 0 \\ \hline
0\,\,\vline &  &  \\
0\,\,\vline & \,\,\,\mathbb{M}_{5} &
\end{array}%
\right) ,  \label{bloque5d}
\end{equation}%
where $\mathbb{M}_{5}$ is the matrix representing the parallelization of the five-dimensional manifold $S^2\times R^3$.

\bigskip

As in the six-dimensional case, if one considers $D=7$, the problem can be
fully understood and this will also be explained in detail below. For extra
dimensions higher than three more patience is required. If we take
four extra dimensions, in addition to the trivial toroidal compactifications of the
last section, we are led to the topologies $S^{1}\times S^{3}$, $%
S^{2}\times S^{2}$ or $S^{4}$. With the exception of the first case, where the
parallelization is provided by a simple extension of the procedure to be
explained below for the $S^{3}$ compactification, the parallelization will
be highly non-trivial because $S^{2}$ and $S^{4}$ do not admit a global
basis. This problem turns out to be even more difficult as the dimension
increases, because the number of possible compactifications increases as $D-4$,
the number of extra dimensions.

\bigskip

\subsection{Complete characterization of $D=7$}

As a working example, let us now consider the various possibilities arising
from the case $D=7$. The first of these was worked with in the last section, where
toroidal compactifications were extensively discussed. Now we will consider
the $7-$dimensional metric
\begin{eqnarray}
&&ds^2=dt^2-a_{1}^2(t)[d\psi^2+\sin^2\psi(d\theta^2+\sin^2\theta
\,d\phi^2)]-\notag \\&&a_{0}^2(t)(dx^2+dy^2+dz^2),  \label{met7D}
\end{eqnarray}%
where we have supposed three additional dimensions spherically compactified.
The manifold topology is $R \times S^3\times R^3$, which is a product of
parallelizable manifolds. Following the scheme \eqref{esquema} we will now have

\begin{equation}  \label{esquema7D}
e^{a}_{\mu}=\left(
\begin{array}{ccccccc}
1\,\vline & 0 & 0 & 0\,\,\vline & 0 & 0 & 0 \\ \hline
0\,\vline &  &  & \,\,\,\,\,\vline &  &  &  \\
0\,\vline &  & \mathbb{M}_{3} & \,\,\,\,\,\vline &  & \mathbb{O} &  \\
0\,\vline &  &  & \,\,\,\,\,\vline &  &  &  \\ \hline
0\,\vline &  &  & \,\,\,\,\,\vline & a_{0} & 0 & 0 \\
0\,\vline &  & \mathbb{O} & \,\,\,\,\,\vline & 0 & a_{0} & 0 \\
0\,\vline &  &  & \,\,\,\,\,\vline & 0 & 0 & a_{0}%
\end{array}
\right),
\end{equation}
where now $\mathbb{M}_{3}$ is the matrix representing the paralellization of
$S^{3}$. A global basis for $\mathcal{T}(S^{3})$ has been obtained in
\cite{Nos3}, so we will just summarize the result here. For $\mathbb{M}_{3}$ we
have
\begin{widetext}
\begin{equation}  \label{campos 3D}
\mathbb{M}_{3}=a_{1}(t)\left(
\begin{array}{ccc}
c(\theta) & -s(\psi)c(\psi)s(\theta) & -s^2(\psi)s^2(\theta) \\
s(\theta)c(\phi) & s(\psi)[s(\psi)s(\phi)+c(\psi)c(\theta)c(\phi)] &
s(\psi)s(\theta)[s(\psi)c(\theta)c(\phi)-c(\psi)s(\phi)] \\
s(\theta)s(\phi) & s(\psi)[c(\psi)c(\theta)s(\phi)-s(\psi)c(\phi)] &
s(\psi)s(\theta)[c(\psi)c(\phi)+s(\psi)c(\theta)s(\phi)]%
\end{array}
\right),
\end{equation}
\end{widetext}
where we have used $s\equiv sin$ and $c\equiv cos$. We can just add here that $\mathbb{M}_{3}$ can be obtained from the \emph{naive} triad $%
a_{1}(t)diag(d\psi,\sin\psi\, d\theta,\sin\psi\,\sin\theta\,d\phi)$ by means
of a local Euler rotation (see \cite{Nos3} for details). It is a simple
exercise to check that the frame \eqref{esquema7D} with $\mathbb{M}_{3}$
given by \eqref{campos 3D} gives rise to the metric \eqref{met7D}.

Having found the proper frame, we can now proceed to compute the motion equations. The vielbein (\ref%
{esquema7D}) leads to the invariant
\begin{equation}
T= -6(H_{0}^2+H_{1}^2+3H_{0}H_{1}-a_{1}^{-2}).  \label{inv7d}
\end{equation}
The initial value equation then results in
\begin{equation}
f+12 f^{\prime }(H_{0}^2+H_{1}^2+3H_{0}H_{1})=16 \pi G \rho.  \label{eqvi7d}
\end{equation}%
On the other hand, by varying respect the spatial sector of the vielbein, we
get two equations:

\begin{eqnarray}
&&2f^{\prime }\left(7 H_{0}^2+4H_{1}^2+9H_{0}H_{1}+3\dot{H}_{0}+2\dot{H}_{1}-%
\frac{T}{3}\right)+\notag\\&&2f^{\prime \prime }\dot{T}(3H_{0}+2H_{1})+f=-16 \pi G
p_{1},  \label{esp7d1}
\end{eqnarray}

\begin{eqnarray}
&&2f^{\prime }\left(6 H_{0}^2+9H_{1}^2+15H_{0}H_{1}+2\dot{H}_{0}+3\dot{H}%
_{1}\right)+\notag\\&&2f^{\prime \prime }\dot{T}(2H_{0}+3H_{1})+f=-16 \pi G p_{0}.
\label{esp7d2}
\end{eqnarray}%
All the remaining equations are trivial or equal to these last two. This is proof that the frame \eqref{esquema7D} is actually a parallelization of \eqref{met7D}, because it leads to a consistent set of motion equations.

\bigskip

It is interesting to compare this compactification with the toroidal ones
(in $D=7$) worked with in the previous section. A remarkable property of the system
\eqref{eqvi7d}-\eqref{esp7d2} is that it admits a de Sitter expansion for the
bulk while the extra dimensions remains compactified with a constant scale
factor, and that this happens for a pure vacuum. Actually, if we set $H_{0}$
as a constant and $H_{1}=\rho=p_{1}=p_{2}=0$, the system becomes (note that $%
T=-6(H_{0}^2-a_{1}^{-2})$ is constant in this case)

\begin{eqnarray}
&&f+12H_{0}^2f^{\prime }=0  \label{eqvi7d2sol} \\
&&2f^{\prime }(7 H_{0}^2-\frac{T}{3})+f=0,  \label{esp7d2sol}
\end{eqnarray}%
equation \eqref{esp7d2} being equal to \eqref{eqvi7d2sol}. Inserting the
expression $T=-6(H_{0}^2-a_{1}^{-2})$ and the value of $f^{\prime }$ from
\eqref{eqvi7d2sol} in \eqref{esp7d2sol} we immediately get
\begin{equation}
H_{0}^2=\frac{2}{3}a_{1}^{-2} , \label{hya}
\end{equation}%
which implies $T=3H_0^2$. Note that $f(T)=T$ is not a solution of equations \eqref{eqvi7d2sol} and \eqref{esp7d2sol}, showing that this kind of solution is absent in Einstein's theory. The result \eqref{hya} is actually valid only for smooth deformations of GR, i.e., for $f(T)$ models of the form $f(T)=T+\beta\, g(T)$. Replacing this form of $f(T)$ in Eq. \eqref{eqvi7d2sol}, we obtain a relationship between $H_0$, $\beta$ and the potential constants appearing in $g(T)$.

Also, note that at this point the constant value of $H_{0}$ (or $a_{1}$)
remains undetermined. Nevertheless, given that the observational evidence
suggests that the size of the extra dimensions, if non-null, should be very
small at the present time, Eq. \eqref{hya} focuses attention on the
ultraviolet regime. This enables us to interpret inflation as a vacuum-driven
expansion given by the small compact extra dimensions. For this purpose let
us invoke, as an example, the high energy deformation of GR with the Born-Infeld
structure worked in \cite{Nos1} and \cite{Nos2}. We thus have

\begin{equation}
f(T)=-\lambda\,[\sqrt{1-2\lambda^{-1}\,T}-1],  \label{deff}
\end{equation}%
where $\lambda$ is the Born-Infeld constant (so, $\beta=-\lambda^{-1}$, $g(T)=T^2/2+\lambda^2\,\mathcal{O}(T/\lambda)^3$). In this manner we have that
Einstein's gravity (in its absolute paralellism form) is recovered when $%
T/\lambda<<1$. The role of Eq. \eqref{eqvi7d2sol} will be to link the
undetermined parameter $H_{0}$ (or $a_{1}$) with $\lambda$. Actually,
replacing the functional form \eqref{deff} in \eqref{eqvi7d2sol} we get the
quadratic expression

\begin{equation}
(3u+1)^2=1+u,\,\,\,\,\,\,\,\,\,u=-6\lambda^{-1}H_{0}^2,  \label{cuadratica}
\end{equation}%
which leads to the solutions $H_{0}=0$ and $H_{0}^2=5 \lambda/54$. Of
course, $H_{0}=0$ just says that Minkowski spacetime is a vacuum solution of
the theory. In turn, $H_{0}^2=5 \lambda/54$, which is very close to $%
H_{0}^2= \lambda/12$, the value obtained in \cite{Nos2} for the inflationary
era, represents a de Sitter accelerated expansion, the cause of which is the
presence of the extra dimensions.

\bigskip

Let us go back now to the full characterization of $D=7$. The remaining topology for the three additional dimensions is just $%
S^{1}\times S^{2}$, so the full metric looks in this case like
\begin{eqnarray}
&&ds^2=dt^2-a_{2}^2(t)d\Theta^2 -a_{1}^2(t)(d\theta^2+\sin^2\theta \
d\phi^2)-\notag\\ &&a_{0}^2(t)(dx^2+dy^2+dz^2).  \label{metD=7otra}
\end{eqnarray}%
We proceed now to find the parallel fields for the geometry (\ref{metD=7otra}%
). For this task we focus on the embedding of the submanifold $\mathcal{M}%
_{3}=S^{1}\times S^{2}$ in the four-dimensional space $S^{1}\times R^3$. Installing coordinates $(\Theta,X,Y,Z)$ in this manifold, we have that a parallelization of $\mathcal{T}^{*}(\mathcal{M}_{3})$ reads

\begin{eqnarray}
E^1&=&a_{2}(t) Z d\Theta-a_{1}(t)(Y dX-X dY)  \notag \\
E^2&=&a_{2}(t) Y d\Theta+a_{1}(t)(Z dX-X dZ)  \notag \\
E^3&=&a_{2}(t) X d\Theta -a_{1}(t)(Z dY -Y dZ).  \label{paralelizacion}
\end{eqnarray}%
This base turns out to be a global parallelization for $S^{1}\times S^{2}$ only if $a_{1}(t)$ and $a_{2}(t)$ are non-null, so a singularity in the Riemannian sense (i.e., when at least either $a_{1}(t)$ or $a_{2}(t)$ vanishes), actually represents a singularity in the parallelization process. In terms of the basis $(\Theta,\theta,\phi)$, where $X=\sin\theta\cos\phi$, $Y=\sin\theta\sin\phi$ and $Z=\cos\theta$ we therefore have

\begin{eqnarray}
E^1&=&a_{2}(t)\cos\theta \,d\Theta+a_{1}(t) \sin^2\theta \,d\phi  \notag \\
E^2&=&a_{2}(t)\sin\theta\sin\phi \,d\Theta+\notag\\ &&a_{1}(t)(\cos\phi \,d\theta-\sin\theta \sin\phi\cos\theta\, d\phi) \notag \\
E^3&=&a_{2}(t)\sin\theta\cos\phi \,d\Theta-\notag\\ &&a_{1}(t)(\sin\phi \,d\theta+\cos\theta\sin\theta\cos\phi\, d\phi).  \label{paralelizacionenbase}
\end{eqnarray}%
Being $S^{1}\times S^{2}$ a three dimensional manifold, we can only provide a glimpse of the structure underlying the fields (\ref{paralelizacion}); in every point of the submanifold $\Theta=constant$, we have
\begin{eqnarray}
\tilde{E}^1&=&-Y dX+X dY  \notag \\
\tilde{E}^2&=&Z dX-X dZ  \notag \\
\tilde{E}^3&=&-Z dY +Y dZ,  \label{paralelizacionsub}
\end{eqnarray}%
where $\tilde{E}^i=E^i/a_{1}(t)$. These fields, as viewed immersed in 3D-euclidian space, look similar to the ones of Figure 1 of Ref. \cite{Nos3}, where the parallelization (\ref{campos 3D}) is showed for $\psi=\pi/2$ (i.e. for the $S^3$ sphere hyper-equator). However, this resemblance is partial, and it is important to bear in mind that the fields (\ref{campos 3D}) and (\ref{paralelizacionenbase}) represent parallelizations of two genuinely different topological structures, $S^3$ and $S^1\times S^2$ respectively.

In analogy with the analysis performed before, we can now obtain the motion equations. First, it is necessary to compute the scalar invariant, which reads
\begin{equation}
T= -2(3H_{0}^2+H_{1}^2+6H_{0}H_{1}+2H_{1}H_{2}+3H_{0}H_{2}-a_{1}^{-2}).
\label{inv7dos}
\end{equation}
Note that just the scale factor corresponding to $S^2$ is present in $T$, as must be the case. Moreover, the scalars  \eqref{inv7d} and \eqref{inv7dos} are different even when $a_{0}=a_{1}$, because they represent different geometries. The full set of motion equations are

\begin{widetext}
\begin{eqnarray}
&&f+4 f^{\prime }\left(H_{1}^{2}+3H_{2}H_{0}+3H_{0}^{2}+2H_{1}H_{2}+6H_{1}H_{0}\right) =16\pi G\rho,\label{valin7dos} \\
&&f+2f^{\prime }\left(4H_{1}^{2}+2H_{1}H_{2}+9H_{0}^{2}+
12H_{1}H_{0}+3H_{0}H_{2}+2\dot{H}_{1}+3\dot{H}_{0}\right)
+2f^{\prime \prime }\left( 2H_{1}+3H_{0}\right) \dot{T}=-16\pi G\,p_{2},\label{esp7dos}\\
&&f+2f^{\prime }(
-a_{1}^{-2}+2H_{1}^{2}+H_{2}^{2}+9H_{0}^{2}+3H_{1}H_{2}+9H_{1}H_{0}+6H_{2}H_{0}+%
\dot{H}_{1}+\dot{H}_{2}+3\dot{H}_{0})+\notag \\ &&2f^{\prime \prime }(H_{1}+H_{2}+3H_{0})\dot{T}=-16\pi G\,p_{1},\label{esp7tres}\\
&&f+2f^{\prime }(4H_{1}^{2}+6H_{0}^{2}+H_{2}^{2}+4H_{1}H_{2}+10H_{1}H_{0}+5H_{2}H_{0}+2\dot{H}%
_{1}+\dot{H}_{2}+2\dot{H}_{0})+\notag \\ &&2f^{\prime \prime }(2H_{1}+H_{2}+2H_{0}) \dot{T}=-16\pi
G\,p_{0}.\label{esp7cuatro}
\end{eqnarray}
\end{widetext}
A simple exercise shows that vacuum solutions with $H_{1}=H_{2}=0$ and a non-null constant $H_{0}$ does not exist at all for this specific compactification. Imposing these conditions on \eqref{valin7dos} and \eqref{esp7dos} we are led to the inconsistent (when $H_{0}\neq0$) equations

\begin{eqnarray}
&&f+12 f' H_{0}^2=0, \notag \\
&&f+18 f' H_{0}^2=0. \label{inconsistent}
\end{eqnarray}
So, as far as vacuum-driven inflation given by the small compact extra dimensions is concerned, we see that the $S^{3}$ compactification is clearly favored.

\subsection{Complete characterization of $D=6$}

In this subsection we consider the remaining topology ($S^{2}$) for the internal dimensions of the six-dimensional manifold. We are now ready to go back to the metric \eqref{metD=6}. The topology of this manifold is now described by $R\times S^{2}\times R^{3}$, so we proceed to parallelize this submanifold using a similar method to that employed at the end of the previous subsection. The key point is that we can unwrap the periodic coordinate $\Theta$ in (\ref{paralelizacionenbase}) and think about it as one of the bulk coordinates, let us say $x$. In this way, we obtain the submanifold $\mathcal{M}_{3}=S^2\times R$. Again, embedding this in order to obtain three-dimensional manifold in the four-dimensional space $R^{3}\times R$ and installing coordinates $(X,Y,Z,x)$ on it, we obtain that a parallelization of $\mathcal{T}^{*}(\mathcal{M}_{3})$ reads

\begin{eqnarray}
E^1&=&a_{0}(t)\cos\theta \,dx+a_{1}(t) \sin^2\theta \,d\phi  \notag \\
E^2&=&a_{0}(t)\sin\theta\sin\phi \,dx+\notag \\
&&a_{1}(t)(\cos\phi \,d\theta-\sin\theta \sin\phi\cos\theta\, d\phi) \notag \\
E^3&=&a_{0}(t)\sin\theta\cos\phi \,dx-\notag \\
&&a_{1}(t)(\sin\phi \,d\theta+\cos\theta\sin\theta\cos\phi\, d\phi).
\end{eqnarray}
where again, $X=\sin\theta\cos\phi$, $Y=\sin\theta\sin\phi$ and $Z=\cos\theta$ was used. An observation is in order: nothing prevents thinking of the unwrapped $\Theta$ periodic coordinate now as the $y$ or $z$ coordinate of the bulk given the isotropy and homogeneity of the latter. This actually means that we obtain three different but equivalent sets of parallel vector fields for \eqref{metD=6}.

With this vielbein in mind we can obtain the following torsion invariant and fields equations,

\begin{equation}
T=-2\left( -a_{1}^{-2}%
+H_{1}^{2}+6H_{0}H_{1}+3H_{0}^{2}\right) \label{caso6segundaT}
\end{equation}
\begin{equation}
f+4f^{\prime }\left( H_{1}^{2}+6H_{0}H_{1}+3H_{0}^{2}\right) =16\pi G\rho \label{caso6segunda}
\end{equation}
\begin{eqnarray}
&&f+2f^{\prime }\left( 6H_{0}^{2}+10H_{0}H_{1}+4H_{1}^{2}+2\dot{H}_{0}+2\dot{H}%
_{1}\right) +\notag \\
&&2f^{\prime \prime }\dot{T}\left( 2H_{0}+2H_{1}\right) =-16\pi Gp_{0} \label{caso6segundatres}
\end{eqnarray}
\begin{eqnarray}
&&f+2f^{\prime }\left(-a_{1}^{-2}%
+9H_{0}^{2}+9H_{0}H_{1}+2H_{1}^{2}+3\dot{H}_{0}+\dot{H}_{1}\right)+\notag \\
&&2f^{\prime \prime }\dot{T}\left( 3H_{0}+H_{1}\right) =-16\pi Gp_{1} \label{caso6ddos}
\end{eqnarray}

Again, we are interested in solutions with $\rho =p_{0}=p_{1}=H_{1}=0$ and constant $H_{0}$. From \eqref{caso6segunda} we immediately get
\begin{equation}
f+12 f' H_{0}^2=0,\label{vinculo}
\end{equation}
which is consistent with (\ref{caso6segundatres}). This expression replaced in (\ref{caso6ddos}) leads to
\begin{equation*}
 H_{0}^2=\frac{1}{3}a_{1}^{-2}.
\end{equation*}
This last equation is the analog of \eqref{hya}. Note that this value of $H_{0}$ yields $T=0$ by using \eqref{caso6segundaT}, which turns out to be problematic. Actually, every ultraviolet deformation looks like $f(T)=T+O(T^2)$, so we have $f(0)=0$, $f'(0)=1$, and then \eqref{vinculo} is satisfied only when $H_{0}=0$. We can therefore conclude that physically admissible models with this topology do not contain $H_{1}=0$ and a non-null constant $H_{0}$ as solution.

\bigskip

\section{Concluding comments}
\label{sec:conclution}

This work was devoted to the study of spacetimes with FRW$^{4} \times \mathcal{M}^{D-4}$ topology in the context of the new gravitational schemes with absolute parallelism known as $f(T)$ theories. We have assumed that FRW$^{4}$ represents the spatially flat Friedmann-Robertson-Walker manifold corresponding to the four-dimensional spacetime of standard cosmology, while $\mathcal{M}^{D-4}$ refers to the $(D-4)$-dimensional compact extra dimensions constituting internal space. On one hand we focussed the analysis on $(D-4)$-dimensional manifolds consisting of $(D-4)$ copies of the torus, and on the other hand, several spherical compactifications were considered.

While obtaining the proper parallel vector fields for the toroidal compactifications realized in section III is trivial, (because the $(D-4)$ torus $T^{D-4}=S^{1}\times...\times S^{1}$ is a product of trivial parallelizable manifolds), the corresponding characterization of the spherical topologies is highly non-trivial. Several aspects of this last issue were discussed in section IV, where the problem was posed in detail. Then, we characterized the six and seven-dimensional cases by finding the proper vielbeins for $\mathcal{M}^{2}=S^2$ and $\mathcal{M}^{3}=S^1\times S^{2},\,\,\mathcal{M}^{3}=S^{3}$ respectively. These vielbeins constitute the starting point for \emph{any} $f(T)$ cosmological model of compact extra dimensions, because they represent the basis responsible for the parallelization of the manifold under consideration, i.e., they define the spacetime structure.

In both sections, we discussed a number of simple exact solutions of the field equations. One of the simple solutions explored throughout the different compactifications considered in this work concerns the existence of a constant scale factor for the compact dimensions while the external scale factor expands exponentially in time. This situation corresponds to a vacuum-driven inflation powered by the presence of extra dimensions. We showed that for five and six spacetime dimensions, this kind of behavior does not exist at all. In turn, this problem finds a solution in $D=7$ but not for any kind of compactification, just for $S^3$. In this last case we found that for smooth deformations of GR, the relation between $H_{0}$ and $a_{1}$ is given by equation $H_{0}^2=\frac{2}{3}a_{1}^{-2}$. Therefore, Hubble rates characterizing the inflationary era find their hugeness in the exceedingly tiny value of $a_{1}$. We do not expect this exact solution to describe all the details in the compactification process, but merely the asymptotic stage of it. We think that a solution for the internal dimensions of the form $a_{in}=a_{1}+ \alpha\, Exp(-H_{0}t)$, with a constant $\alpha$ certainly do exist in theory, though it will probably require the presence of matter fields and the use of numerical techniques.

However, the existence of the exact solution in $D=7$ with $S^3$ topology for the extra dimensions leads us to think that a similar exact solution will exist \emph{only} in $D=11$, where the seven additional dimensions will be compactified with topology $S^7$. Note that $S^7$ is the only remaining parallelizable sphere, so vielbein fields with product structure FRW$^{4}\times S^{7}$ will constitute a parallelization of the entire manifold. In this respect, we proved in section~\ref{sec:toroidal} that toroidal compactifications certainly do not lead to this kind of solution in any spacetime dimension $D$. It will be a matter for future investigations to find the parallel vector fields of FRW$^{4}\times S^{7}$ and those of the other admissible eleven-dimensional spherical topologies, and confirm (or dismiss) this conjecture. The veracity of this statement may make it possible to detect interesting links between these $f(T)$ deformed gravitational schemes and the low energy limit of M-theory.

\bigskip

Much more work is certainly necessary to fully understand the peculiarities present in the parallelization process behind $f(T)$ theories. In particular, the characterization of all the spherical topologies in a cosmological context for a given spacetime dimension (with $D>7$) remains entirely open. We would like to think that our work constitutes a first and significant step towards that goal.

\acknowledgments
F.F. is indebted to F. Canfora for many enlightening discussions about
extra dimensions in cosmology. F.F. is supported by CONICET and Universidad
de R\'{i}o Negro. Y. V. is supported by FONDECYT grant 11121148, and by Direcci\'{o}n de Investigaci\'{o}n y Desarrollo, Universidad de la Frontera, DIUFRO
DI11-0071. P.G. and Y.V. acknowledge the hospitality of the Centro At\'{o}mico
Bariloche where part of this work was undertaken.


\begin{thebibliography}{99}


\bibitem{ein28} A.~Unzicker and T.~Case, arXiv:0503046.

\bibitem{Hayashi79} K. Hayashi and T. Shirafuji, \emph{Phys. Rev.} {\bf D19}, (1979) 3524,  \emph{Addendum-ibid}.
 {\bf D24}, (1982) 3312.

\bibitem{Nos1} R.~Ferraro and F.~Fiorini, \emph{Phys.\ Rev.} {\bf D75}
(2007) 084031.

\bibitem{Nos2} R.~Ferraro, F.~Fiorini, \emph{Phys.\ Rev.} {\bf D78}, (2008) 124019.

\bibitem{Bengochea:2008gz} G.~R.~Bengochea and R.~Ferraro, \emph{Phys.\ Rev.} {\bf D79}, (2009) 124019.

\bibitem{Linder:2010py} E.~V.~Linder, \emph{Phys.\ Rev.} {\bf D81},
(2010) 127301.


\bibitem{Zheng:2010am} R.~Zheng, Q.~-G.~Huang, \emph{JCAP} {\bf1103}, (2011) 002.


\bibitem{Bamba:2010wb} K.~Bamba, C.~-Q.~Geng, C.~-C.~Lee, L.~-W.~Luo, \emph{JCAP} {\bf
1101}, (2011) 021.

\bibitem{Zhang:2011qp} Y.~Zhang, H.~Li, Y.~Gong, Z.~-H.~Zhu, \emph{JCAP} {\bf1107}, (2011) 015.

\bibitem{Capozziello006} S.~Capozziello, V.~F.~Cardone, H.~Farajollahi,
A.~Ravanpak, \emph{Phys.\ Rev.} {\bf D84}, (2011) 043527.


\bibitem{Bamba:2011pz} K.~Bamba, C.~-Q.~Geng, \emph{JCAP} {\bf 1111}, (2011) 008.


\bibitem{Wei:2011yr} H.~Wei,
\emph{Phys.\ Lett.} {\bf B712}, (2012) 430.


\bibitem{Geng:2011ka} C.~-Q.~Geng, C.~-C.~Lee, E.~N.~Saridakis, \emph{JCAP} {\bf 1201}, (2012) 002.


\bibitem{Karami:2012fu} K.~Karami and A.~Abdolmaleki, \emph{JCAP} {\bf 1204} (2012)
007.

\bibitem{Yang:2012hu} J.~Yang, Y.~-L.~Li, Y.~Zhong and Y.~Li, \emph{Phys.\ Rev.} {\bf  D85},
(2012) 084033.

\bibitem{Xu:2012jf} C.~Xu, E.~N.~Saridakis and G.~Leon, \emph{JCAP} {\bf 1207}, (2012) 005.

\bibitem{Bamba:2012vg} K.~Bamba, R. Myrzakulov, S. Nojiri and
S.~D.~Odintsov, \emph{Phys.\ Rev.} {\bf
D85} (2012) 104036.

\bibitem{Ghosh:2012pg} R.~Ghosh, A. Pasqua and S.~Chattopadhyay, \emph{Eur. Phys. J. Plus} (2013) 128:12.

\bibitem{review}
S. Capozziello and M. De Laurentis,
\emph{Phys. Rept.} {\bf 509},   (2011) 167.

\bibitem{Kaluza:1921tu}
  T.~Kaluza,
  \emph{Sitzungsber.\ Preuss.\ Akad.\ Wiss.\ Berlin (Math.\ Phys.)} {\bf 1921}, (1921) 966.

\bibitem{Font:2005td}
  A.~Font and S.~Theisen,
  \emph{Lect.\ Notes Phys.}  {\bf 668}, (2005) 101.

\bibitem{Chan:2000ms}
  C.~S.~Chan, P.~L.~Paul and H.~L.~Verlinde,
  \emph{Nucl.\ Phys.} {\bf B581},  (2000) 156.


\bibitem{Chodos:1979vk}
  A.~Chodos and S.~L.~Detweiler,
  \emph{Phys.\ Rev. } {\bf D21}, (1980) 2167.

\bibitem{Freund:1982pg}
  P.~G.~O.~Freund,
  \emph{Nucl.\ Phys. } {\bf B209},  (1982) 146.

\bibitem{Dereli:1982ar}
  T.~Dereli and R.~W.~Tucker,
  \emph{Phys.\ Lett. } {\bf B125},  (1983) 133.


\bibitem{Demianski:1987rm}
  M.~Demianski, Z.~Golda, L.~M.~Sokolowski, M.~Szydlowski and P.~Turkowski,
 \emph{ J.\ Math.\ Phys.}  {\bf 28}, (1987) 171 .

\bibitem{Szydlowski}
  M.~Biesiada, M.~Szydlowski and J.~Szczesny,
  \emph{Acta Phys.\ Polon.} {\bf B19}, (1988) 3.

\bibitem{Szydlowski:1990jt}
  M.~Szydlowski, A.~Lapeta and B.~Lapeta,
\emph{Acta Phys.\ Polon.} {\bf B21}, (1990) 627 .

\bibitem{Szydlowski:1990ju}
  M.~Szydlowski, A.~Lapeta and B.~Lapeta,
 \emph{ Acta Phys.\ Polon. } {\bf B21},  (1990) 643.

 \bibitem{emparan:2013}
  R. Emparan, R. Suzuki and K. Tanabe, \emph{JHEP} \textbf{1306} (2013) 009.

\bibitem{Gaston:2013}
  G. Giribet, \emph{Phys. Rev.} \textbf{D87} (2013) 107504.

\bibitem{Capozziello:2012zj}
  S.~Capozziello, P.~A.~Gonzalez, E.~N.~Saridakis and Y.~Vasquez,
   \emph{JHEP} {\bf 1302} (2013) 039.

\bibitem{Odintsov2} K. Bamba, S. Nojiri and S. D. Odintsov, \emph{Phys. Lett.} {\bf B725} (2013) 368.

\bibitem{maartens:2010}
  R. Maartens and K. Koyama,
   \emph{Living. Rev. Rel.} {\bf 13} (2010) 5.


\bibitem{Nos3} R. Ferraro and F. Fiorini, \emph{Phys. Lett.} {\bf B702} (2011) 75. See
also R. Ferraro and F. Fiorini, \emph{IJMP} (Conference Series), {\bf 3} (2011) 227.

\bibitem{Nos4} R. Ferraro and F. Fiorini, \emph{Phys. Rev.} {\bf D84} (2011) 083518.

\bibitem{Izumi} K. Izumi and Y. C. Ong, \emph{JCAP} \textbf{1306} (2013) 029.

\bibitem{Barrow} B. Li, T. P. Sotiriou and J. D. Barrow, \emph{Phys. Rev.} {\bf D83} (2011) 104017.

\bibitem{Youssef} N. L. Youssef and W. A. Elsayed, arXiv:1209.1379 (gr-qc).

\bibitem{Hehl} F. W. Hehl, J. D. McCrea, E. W. Mielke and Y. Ne'eman, \emph{Phys. Rep.} {258}
(1995) 1.

\bibitem{christian} N. Tamanini and C. G. Boehmer, Phys. Rev. \textbf{D86} (2012) 044009.

\bibitem{Miao:2011} M. Li, R. X. Miao and Y. G. Miao, \emph{JHEP} {\bf 1107}  (2011) 108.

\bibitem{Sti} E. Stiefel, \emph{Comm. Math. Helv.} {\bf 8} (1936) 305.

\bibitem{Ker} M. A. Kervaire and J. W. Milnor, \emph{Annals of Math.} {\bf 77} (1963) 504.

\bibitem{Hai} M. Eisenberg and R. Guy, \emph{Am. Math. Mon.} {\bf 86}, 7 (1979) 571.

\end{thebibliography}
\end{document}